\documentclass[acmsmall]{acmart}
\renewcommand\footnotetextcopyrightpermission[1]{}
\usepackage{multirow}
\usepackage{makecell}
\usepackage{longtable}
\usepackage{tabularx}
\usepackage{tikz}
\usetikzlibrary{positioning,chains}
\usepackage[table]{xcolor}

\usepackage[final]{changes}

\AtBeginDocument{%
  \providecommand\BibTeX{{%
    \normalfont B\kern-0.5em{\scshape i\kern-0.25em b}\kern-0.8em\TeX}}}

\setcopyright{none}
\copyrightyear{2021}
\acmYear{2021}
\acmDOI{}

\acmConference{To Appear in PACM Human-Computer Interaction}
\acmBooktitle{CSCW '21}
\acmPrice{}
\acmISBN{}

\begin{document}

\title[Problematic Machine Behavior]{Problematic Machine Behavior: A Systematic Literature Review of Algorithm Audits}

\author{Jack Bandy}
\affiliation{%
  \institution{Northwestern University}
  \city{Evanston}
  \state{Illinois}
  \country{USA}
}
\email{jackbandy@u.northwestern.edu}

\begin{abstract}
While algorithm audits are growing rapidly in commonality and public importance, relatively little scholarly work has gone toward synthesizing prior work and strategizing future research in the area. This systematic literature review aims to do just that, following PRISMA guidelines in a review of over 500 English articles that yielded 62 algorithm audit studies. The studies are synthesized and organized primarily by behavior (discrimination, distortion, exploitation, and misjudgement), with codes also provided for domain (e.g. search, vision, advertising, etc.), organization (e.g. Google, Facebook, Amazon, etc.), and audit method (e.g. sock puppet, direct scrape, crowdsourcing, etc.). The review shows how previous audit studies have exposed public-facing algorithms exhibiting problematic behavior, such as search algorithms culpable of distortion and advertising algorithms culpable of discrimination. Based on the studies reviewed, it also suggests some behaviors (e.g. discrimination on the basis of intersectional identities), domains (e.g. advertising algorithms), methods (e.g. code auditing), and organizations (e.g. Twitter, TikTok, LinkedIn) that call for future audit attention. The paper concludes by offering the common ingredients of successful audits, and discussing algorithm auditing in the context of broader research working toward algorithmic justice.
\end{abstract}


\maketitle

\section{Introduction}
Algorithm auditing has become an important research method for \added{diagnosing problematic behavior in algorithmic systems}. For example, do targeted advertising algorithms facilitate discrimination? \cite{Sweeney2013,Lambrecht2019a} Does YouTube's recommendation algorithm elevate extremist videos? \cite{Faddoul} Do facial recognition algorithms perform worse on \added{darker-skinned} females? \cite{Raji} Does Google's search algorithm favor certain news outlets? \cite{Robertson2018c,Trielli} \added{These questions continue to multiply as algorithmic systems become more pervasive and powerful in society.} 

Despite the growing commonality and public importance of algorithm audits that address these problems, relatively little work has gone toward clarifying the past trajectory and future agenda of algorithm auditing. Systematic literature reviews (SLRs) -- sometimes referred to as literature surveys \cite{Wobbrock2016} -- have been a prominent part of computing research \cite{Froehlich2010,Dell,Lopez2017,Abdul2018,Harris2019}, and help "identify trends and gaps in the literature" \cite{Dillahunt2017} to clarify shared research goals for future work. To this end, the present paper conducts a \added{scoped} SLR of algorithm audits, screening over 500 papers from a variety of journals and conferences. By thematically analyzing 62 studies that audited public-facing algorithmic systems, we identify a taxonomy of problematic machine behaviors that helps organize prior work as well as strategize for future work.

Based on the review, we identify four major types of problematic behavior in algorithmic systems: discrimination, distortion, exploitation, and misjudgement. Out of the 62 studies reviewed, most focused on discrimination (N=21) or distortion (N=29). Audit studies also gave more attention to search algorithms (N=25), advertising algorithms (N=12), and recommendation algorithms (N=8), helping to \added{diagnose} a range of problematic behaviors on these systems. Taken together, \added{these empirical studies provide substantial evidence that the public harms of algorithmic systems already exist in real-world systems, not merely in hypothetical scenarios}. These previous studies also help chart a promising path forward for future audits that can serve as a diagnostic for "how social problems manifest in technical systems" \cite{Abebe2019}.

\added{Specifically, the review suggests that some behaviors, domains, and methods call for future audit attention. In terms of behaviors, future audits should further explore discrimination on the basis of intersectional identity. For example, dynamic pricing algorithms might be audited to directly measure the allocative harms of price discrimination on the basis of intersectional identities pertaining to race, age, sex, gender, and more. This paper joins a growing chorus of calls (e.g. \cite{Buolamwini2018,PenaGangadharan2019,Hoffmann2019,d2020datafeminism,Hanna2020,Ogbonnaya-Ogburu2020}) for computing research to recognize identity as multi-faceted rather than one-dimensional. In terms of domains that deserve further research attention, advertising algorithms form the economic backbone of large technology companies, and they pose a number of potential allocative and representational harms to the public. Code auditing appears to be under-explored as an audit method, even though a growing number of influential algorithms are open-sourced. Finally, the studies we reviewed offered limited insights on some organizations such as Twitter, LinkedIn, and TikTok, even though these organizations operate widely influential algorithmic systems.}

Formally, this literature review reports what previous algorithm audits have done (RQ1) and what remains to be done in future audits (RQ2). It also outlines important ingredients for successful audit studies, then concludes by discussing algorithm auditing as \added{a diagnostic tool} within a broader research effort working toward algorithmic justice.


\section{Related Work}
Literature reviews (sometimes referred to as literature surveys \cite{Wobbrock2016}) have been a prominent part of computing research \cite{Froehlich2010,Dell,Lopez2017,Abdul2018,Harris2019}, and a number of reviews are closely related to this work. A 2017 book by Cathy O'Neil \cite{ONeil} illustrated how various algorithmic systems create disparate impact, and the chapters discuss \added{discrimination} in hiring algorithms, loan approval algorithms, labor-scheduling algorithms, and more. \citet{Sandvig2014} review methods for algorithm auditing, and include some examples of each method. A 2016 review article sought to "map the debate" around ethics and algorithms \cite{Mittelstadt}, and a 2018 review article \cite{Abdul2018} identified a research agenda for developing "Explainable, Accountable and Intelligible Systems." While they do overlap with the present work, these studies did not address algorithm auditing as a primary topic.

One closely-related literature review examined ethical considerations for data science \cite{Saltz2019}. The authors identified three challenges related to data (privacy/anonymity, misuse, and accuracy/validity), as well as three challenges related to mathematical modeling (personal/group harm, subjective model design, and model misuse/misinterpretation). Since these challenges are pertinent to both data science and algorithmic systems, many algorithm audits center around the same topics, especially personal/group harm.

\added{There has also been important related work in developing frameworks and theories for algorithm auditing and accountability. \citet{Raji2020} introduce a framework intended for internal auditing, which organizations could use when developing algorithms. This framework is exceedingly helpful, although the focus on internal development leads to different audit considerations compared to this review with its focus on public-facing algorithms that have already been deployed.} One related literature review by \citet{Wieringa2020} outlines a theory of "algorithmic accountability" through a synthesis of over 200 English articles. The study uses a conceptual framework for accountability to organize the findings, and concludes with a robust definition for algorithmic accountability: "Algorithmic accountability concerns a networked account for a socio-technical algorithmic system, following the various stages of the system’s lifecycle." Building on this review, the current work focuses specifically on audit studies aimed at algorithmic accountability, and synthesizes the types of behavior that require accountability.

Finally, a 2017 literature review of "The sharing economy in computing" \cite{Dillahunt2017} overlaps with this work by including some algorithm audits of systems such as Uber and TaskRabbit. They labeled papers as "algorithmic auditing" if they described "how algorithms can rule the sharing economy sites," finding nine papers that met this definition in the sharing economy literature. This review expands the scope beyond just the sharing economy, including audits of search algorithms like Google, recommendation algorithms such as those used by Spotify, computer vision algorithms such as Amazon's "Rekognition," and more. Still, the literature review presented by \citet{Dillahunt2017} was instructive for identifying potentially relevant studies. It also posed two clear research questions for SLRs in the field of Human-Computer Interaction, which were used in other SLRs \cite{Quinn2011,Harris2019} and also served as a model for this review's research questions. The first question is about \textit{what has been done} and the second is about \textit{what is next to do}:
\begin{itemize}
\item {\textbf{RQ1}}: What kinds of problematic machine behavior have been \added{diagnosed} by previous algorithm audits?
\item {\textbf{RQ2}}: What remains for future algorithm audits to examine the problematic ways that algorithms exercise power in society?
\end{itemize}

\section{Methods}
To answer these two research questions, I conducted a scoped review using the Preferred Reporting Items for Systematic Reviews and Meta-Analyses (PRISMA) guidelines \cite{Moher2009}. A scoping review is useful "when what is needed is not detailed answers to specific questions but rather an overview of a broad field" \cite{Moher2015}, and our broad research question aligned with this goal. PRISMA was originally developed for health sciences to systematically asses interventions through meta-analyses, and its evidence-based standards for transparency, comprehensiveness, and bias mitigation have led to usage in a number of other fields. This includes CSCW literature \cite{Lopez2017,Harris2019} and other computing research \cite{VanLaar2017,Wieringa2020}.

The first two main stages of PRISMA are \textit{identification} -- finding a pool of potentially relevant studies -- and \textit{screening} -- manually reviewing article metadata for potential relevance. These first two steps required a definition of an algorithm audit, before moving to the third and fourth stages of \textit{eligibility} -- full-text review -- and \textit{inclusion} -- the final analysis and synthesis stage.

\subsection{What is an algorithm audit?}
\subsubsection{Social Change and Public Impact} While algorithm auditing can serve many purposes, this literature review focused on audits that provide public accountability as a potential means of positive social change. \citet{Abebe2019} recently suggested that one of the major ways computing can play a role in social change is through diagnosing "how [social problems] manifest in technical systems." In this same spirit of social change, \citet{Raji} offered the following definition of an algorithm audit:
\begin{quote}
"An algorithmic audit involves the collection and analysis of outcomes from a fixed algorithm or defined model within a system. Through the stimulation of a mock user population, these audits can uncover problematic patterns in models of interest."
\end{quote}

Social change and public impact have also become important topics not just for auditing researchers, but for computing researchers at large \cite{Fox2017}. Rather than "assume that [computing] research will have a net positive impact on the world," \cite{Armbrust2018}, scholars are now grappling with the larger societal implications of technology, especially "the ways in which power, extraction, and oppression permeate socio-technical systems" \cite{Fox2017}. Because of this growing focus on social change, our definition of an algorithm audit focuses on the potential for algorithm audits to provide meaningful accountability to the public.

\subsubsection{Definition} After reviewing proposed definitions in the auditing literature, the definition used for the review was \textit{an empirical study investigating a public algorithmic system for potential problematic behavior}. Specifically, this entailed the following:
\begin{itemize}
\item An \textit{empirical study} includes an experiment or analysis (quantitative or qualitative) that generates evidence-based claims with well-defined outcome metrics. It must not be purely an opinion/position paper, although position papers with substantial empirical components were included.
\item An \textit{algorithmic system} is any socio-technical system influenced by at least one algorithm. This includes systems that may rely on human judgement and/or other non-algorithmic components, as long as they include at least one algorithm.
\item A \textit{public} algorithmic system is one used in a commercial context or other public setting such as law enforcement, education, criminal justice, or public transportation.
\item \textit{Problematic behavior} in this study refers to discrimination, distortion, exploitation, or misjudgement, as well as various types of behaviors within each of these categories. A behavior is problematic when it causes harm (or potential harm). In the ACM Code of Ethics, examples of harm include "unjustified physical or mental injury, unjustified destruction or disclosure of information, and unjustified damage to property, reputation, and the environment."\footnote{\url{https://www.acm.org/code-of-ethics}} See \citet{Rahwan} for a discussion of "machine behavior" which guided the notion of "\textit{problematic} machine behavior."
\end{itemize}

\subsection{Identification}
\subsubsection{Keyword Search}
The first author designed a keyword search to identify a wide body of records that were potentially relevant to the initial definition of algorithm audits. Keywords were generated iteratively through exploratory searches on Google Scholar. Early searches, based on initial domain knowledge, included "algorithm auditing," "platform audit," and "algorithmic bias." After inspecting results for these keyword searches, several keywords and keyphrases were added. Due to the empirical component of the definition, a boolean key was added to search for papers that included an analysis, experiment, a study, or an audit. 

Due to its interdisciplinary database, flexible search options, and reproducible results, the Scopus database \cite{ELSEVIER2016} was used to identify records. Given the various disciplines relevant to algorithm auditing, it was important to search records in fields like economics, journalism, and law, rather than only computing. Scopus is known to have an expansive database \cite{Falagas2008}, and other literature reviews in computing have utilized it during the identification phase \cite{Abdul2018,Mittelstadt}.

Scopus' flexible search options also allowed an expanded search by identifying papers that reference any of three influential papers in algorithm auditing. The first author selected these papers after noting they were frequently cited as motivation in relevant studies, and confirming each paper was a highly-cited contribution to the field of algorithm auditing. \citet{Sandvig2014} (300 citations) provides the first methodological overview of algorithm auditing, \citet{Gillespie2014} (1,500 citations) discusses the "relevance of algorithms" as having significant societal impact, and \citet{Kitchin2016} (500 citations) discusses both the relevance of algorithms to society and potential methods to critically examine them (citation counts based on Google Scholar in August 2020). The final boolean search string (Table \ref{table-boolean}) generated 506 initial records.
\begin{table}[]
\resizebox{\textwidth}{!}{
\begin{tabular}{|l|l|ll}
\hline
\multicolumn{2}{|l|}{\textbf{Relevant to algorithm auditing}}                           & \multicolumn{2}{l|}{\textbf{Empirical Study}}                                             \\ \hline
Title, abstract, or keyword contains: & "algorithmic bias"                & \multicolumn{1}{l|}{Title, abstract, or keyword contains:} & \multicolumn{1}{l|}{"study"}      \\ \hline
OR Title, abstract, or keyword contains: & "algorithmic discrimination"      & \multicolumn{1}{l|}{OR Title, abstract, or keyword contains:} & \multicolumn{1}{l|}{"experiment"} \\ \hline
OR Title, abstract, or keyword contains: & "algorithmic fairness"            & \multicolumn{1}{l|}{OR Title, abstract, or keyword contains:} & \multicolumn{1}{l|}{"audit"}      \\ \hline
OR Title, abstract, or keyword contains: & "algorithmic accountability"      & \multicolumn{1}{l|}{OR Title, abstract, or keyword contains:} & \multicolumn{1}{l|}{"analysis"}   \\ \hline
OR Full text contains:                       & "algorithm audit*"                 & \multicolumn{2}{l}{\multirow{5}{*}{}}                                                   \\ \cline{1-2}
OR Full text contains:                       & "algorithmic audit*"              & \multicolumn{2}{l}{}                                                                    \\ \cline{1-2}
OR The paper references:                          & "Auditing Algorithms: Research Methods..." \cite{Sandvig2014}       & \multicolumn{2}{l}{}                                                                    \\ \cline{1-2}
OR The paper references:                          & "Thinking critically about and researching algorithms" \cite{Kitchin2016}       & \multicolumn{2}{l}{}                                                                    \\ \cline{1-2}
OR The paper references:                          & "The Relevance of Algorithms" \cite{Gillespie2014} & \multicolumn{2}{l}{}                                                                    \\ \cline{1-2}
\end{tabular}
}
\caption{The boolean search used to identify potential articles in the Scopus database}
\label{table-boolean}
\end{table}

\subsubsection{Additional Sources}
To improve coverage of the review, additional sources were identified and included throughout the writing process, including citations encountered during full-text screening and during the peer review process. The first author maintained a list of papers that were referenced as algorithm audits during full text screening, and added them to the review pipeline. Some papers were also added to the review pipeline during peer review, when reviewers recommended additional algorithm audit studies. Including these additional studies helps minimize any bias introduced in the keyword search, for example, \added{many papers identified from these sources were published in or before 2015. This suggests the initial Scopus search may have suffered from a recency bias which excluded many studies prior to 2016.} In total, additional sources identified 36 studies as potentially relevant to the review. As with the keyword search, the papers were screened and filtered based on titles and abstracts.

\subsection{Title and Abstract Screening}
During the initial screening, the first author examined the title and abstracts for 503 papers (3 duplicates were removed from the original set of 506). Based on this review, 416 papers did not fit the initial definition, which filtered the set down to 87 potential studies that met or likely met the criteria. 36 additional papers were identified through additional sources, so in total, 123 papers were reviewed in full text screening.

\subsection{Full Text Screening}

\subsubsection{Exclusion Reasons}
The first author screened each of the texts that were potentially relevant and excluded papers that did not match the initial definition of an algorithm audit. While many studies were relevant and important for the research area, many did not constitute an algorithm audit for a variety of reasons:
\begin{itemize}
\item \textbf{Theory or methods} (n=16): the paper focused on theoretical or methodological topics, including proposed fairness metrics and other potential solutions to mitigate problematic behavior. For example, \citet{Badami} propose a method for mitigating polarization in recommender systems, but do not conduct an algorithm audit.
\item \textbf{Non-public} (n=13): the study focused on an algorithmic system that does not directly face the public. For example, \citet{Moller2018a} simulate hypothetical recommendation algorithms to audit for echo chamber effects, rather than auditing a real-world algorithm.
\item \textbf{Non-algorithm} (n=12): the study audited an aspect that may \textit{pertain} to an algorithmic system, such as an interface or input data, but did not substantially audit the algorithm. For example, \citet{May2019a} investigate gender disparities in participation on Stack Overflow, but do not explore how Stack Overflow's algorithms may be implicated in the disparities.
\item \textbf{User study} (n=9): the paper focused on how people experience or perceive an algorithmic system, but did not audit the algorithm (i.e. it was a study of human behavior, not machine behavior). For example, \citet{Diaz2019b} conducted an interview study about an algorithm that calculated neighborhood "walkability" scores, but did not conduct an algorithm audit by testing inputs and outputs.
\item \textbf{Development history} (n=5): the study examined the development, design, and/or implementation of an algorithmic system, but did not substantially audit the algorithm. For example, \citet{DeVito2017} provides insight about Facebook's News Feed algorithm by exploring the history and context of its development.
\end{itemize}

\begin{figure}[h]
\resizebox{0.99 \textwidth}{!}{
\begin{tikzpicture}[
    node distance=15mm and 10mm,
    start chain=going below,
 mynode/.style = {
        draw, rectangle, align=center, text width=5cm,
        font=\small, inner sep=3ex, outer sep=0pt,
        on chain},
mylabel/.style = {
        draw, rectangle, align=center, rounded corners, 
        font=\small\bfseries, inner sep=2ex, outer sep=0pt,
        fill=cyan!30, minimum height=32mm,
        on chain},
every join/.style = arrow,
     arrow/.style = {very thick,-stealth}
                    ] 
\coordinate (tc);
\node[above=of tc,font=\bfseries] {PRISMA Flow Diagram};
\node (n1a) [mynode, left=of tc]    {506 records identified 
                                        through database searching};
\node (n1b) [mynode,right=of tc]    {36 additional records identified\\
                                        \added{from additional sources}};
\node (n2)  [mynode, below=of tc]   {503 records after \\ duplicates removed};
\node (n3)  [mynode,join]   {503 titles and \\ abstracts screened
                                        };
\node (n4)  [mynode,join]   {123 (87+36) full-text articles \\ accessed 
                                            for eligibility};
\node (n5)  [mynode,join]   {62 studies included in review};
\node (n6)  [mynode,join]   {62 studies included analysis};
\node (n3r) [mynode,right=of n3]    {416 records excluded};
\node (n4r) [mynode,right=of n4]    {61 full-text articles \\ 
excluded, with reasons};
\draw[arrow] ([xshift=+22mm] n1a.south) coordinate (a)
                                       -- (a |- n2.north);
\draw[arrow] ([xshift=-22mm] n1b.south) coordinate (b)
                                       -- (b |- n4.north);
\draw[arrow] (n3) -- (n3r);
\draw[arrow] (n4) -- (n4r);
    \begin{scope}[node distance=7mm]
\node[mylabel,below left=-3mm and 11mm of n1a.north west]
                {\rotatebox{90}{Identification}};
\node[mylabel]  {\rotatebox{90}{Screening}};
\node[mylabel]  {\rotatebox{90}{Eligibility}};
\node[mylabel]  {\rotatebox{90}{Included}};
    \end{scope}
\end{tikzpicture}}
    \caption{PRISMA Diagram}
    \label{fig:prisma-flow}
\end{figure}
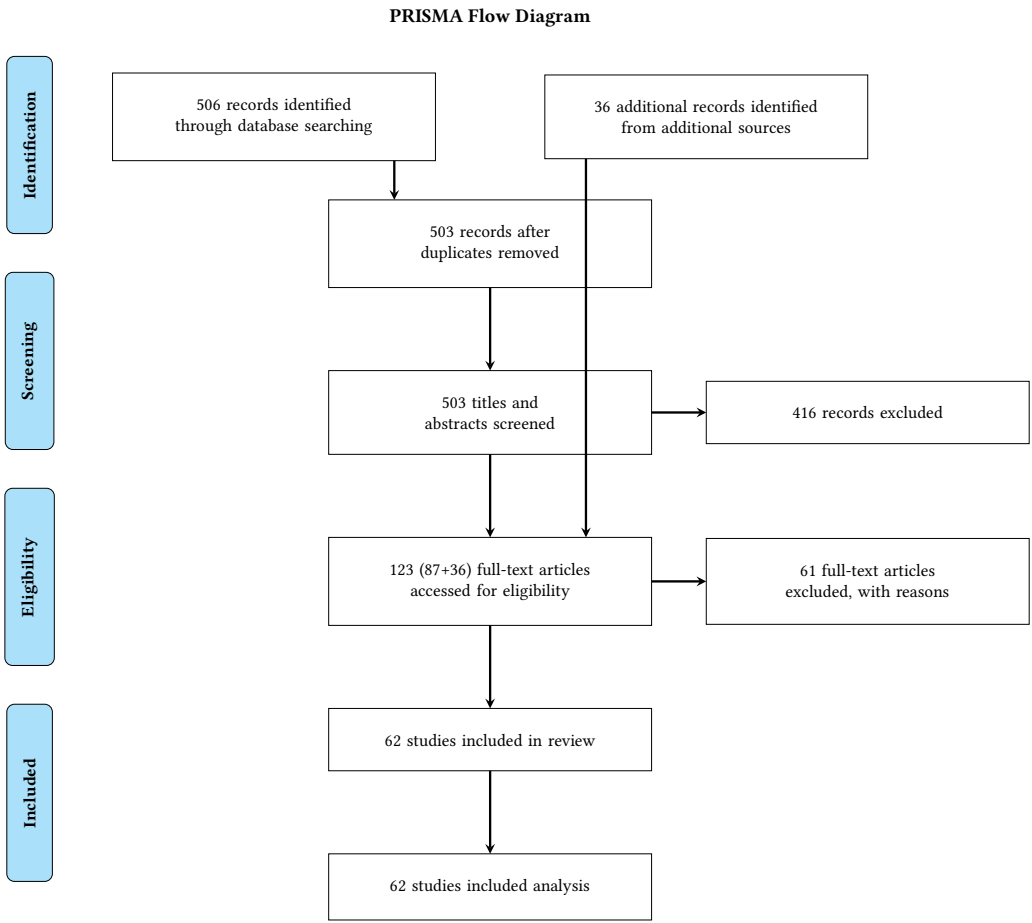

Furthermore, 6 papers could not be reviewed due to lack of access to the full text, resulting in 62 total papers included in the final review. Figure \ref{fig:prisma-flow} shows the flowchart of exclusion and inclusion in terms of the PRISMA guidelines.

\subsection{Analysis and Synthesis}

With 62 relevant studies identified, the author read, summarized and thematically analyzed the papers. Based on categories from two meta-analyses of CSCW scholarship, the first author coded for general data categories: year, domain (e.g. search, vision, recommendation, etc.), and organization audited (e.g. Google, YouTube, Uber, etc.). The first author also coded the audit method used, using the following definitions for audit methods from \citet{Sandvig2014}:
\begin{itemize}
\item \textbf{Code audit}: researchers obtain and analyze the code that makes up the algorithm
\item \textbf{Direct scrape}: researchers collect data directly via an API or other systematic query
\item \textbf{Sock puppet}: researchers collect data by creating computer programs which impersonate users who then test the algorithm
\item \textbf{Carrier puppet}: similar to the sock puppet method, except that the impersonated users affect the real-world system and may "carry" effects onto end users
\item \textbf{Crowdsourcing}: researchers collect data by hiring end users to test the algorithm
\end{itemize}

Finally, to identify themes from the data, the author worked with another coder using inductive thematic analysis \cite{Braun2006a}, which aims to link themes directly to the data. In our case, given the focus on problematic behavior, this meant linking problematic machine behaviors directly to algorithm audit studies. As in prior CSCW literature reviews (e.g. \cite{Harris2019}), one author developed an open and iterative coding scheme using language directly from the included papers. When a study explicitly named a problematic behavior, the phrase was incorporated verbatim into the coding process. The author then iteratively grouped, synthesized, and renamed these themes until the process produced a final scheme of behaviors which worked "in relation to the coded extracts and the entire data set" \cite{Braun2006a}.

After the initial behaviors were defined and named, the second coder used the scheme to independently code 12 random studies (roughly a 20\% sample) from the full set of 62. Inter-rater reliability was moderate in this first round (Cohen's $\kappa=0.42$), after which the reviewers discussed discrepancies and potential modifications to the coding scheme. The first author then coded all papers with a modified coding scheme, and the second coder applied this scheme to another set of 12 random studies. In this second round with the revised scheme, inter-rater reliability was perfect (Cohen's $\kappa=1.0$), and no further revisions were made to the coding scheme for behaviors.

The final scheme included four main types of problematic behavior, each with some variations which are discussed in the results section. Generally, coders found that each of the audits \added{primarily} addressed one of four major types of problematic behavior:

\begin{itemize}
\item \textbf{Discrimination}: the algorithm disparately treats or disparately impacts people \added{on the basis of their} race, age, gender, location, socioeconomic status, and/or intersectional identity. For example, an algorithm implicated in discrimination may systematically favor people who identify as males, or reinforce harmful stereotypes about elderly people.

\item \textbf{Distortion}: the algorithm presents media that distorts or obscures an underlying reality. For example, an algorithm implicated in distortion may favor content from a given political perspective, hyper-personalize output for different users, change its output frequently and without good reason, or provide misleading information to users.

\item \textbf{Exploitation}: the algorithm inappropriately uses content from other sources and/or sensitive personal information from people. For example, an algorithm implicated in exploitation may infer sensitive personal information from users without proper consent, or feature content from an outside source without attribution.

\item \textbf{Misjudgement}: the algorithm makes incorrect predictions or classifications. Notably, misjudgement can often lead to discrimination, distortion, and/or exploitation, but some studies in the review focused on this initial error of misjudgement without exploring second-order problematic effects. An algorithm implicated in misjudgement may incorrectly classify a user’s employment status or mislabel a piece of political news as being primarily about sports, for example.
\end{itemize}

\added{Studies that addressed more than one of the above behaviors were coded based on their primary focus.} The 62 audit studies and their codes are available in a public repository\footnote{https://github.com/comp-journalism/list-of-algorithm-audits/} and will be updated as an effort to account for overlooked studies, as well as studies that were published after this review.

\section{Results for RQ1: Previous Algorithm Audits}

This section provides an overview of the findings for RQ1, "what kinds of problematic machine behavior have been exposed by previous algorithm audits?" It organizes results according to behavior:
\begin{itemize}
\item Discrimination (Section \ref{sec-discrimination})
\item Distortion (Section \ref{sec-distortion})
\item Exploitation (Section \ref{sec-exploitation})
\item Misjudgement (Section \ref{sec-misjudgement})
\end{itemize}

\subsection{Overview}

Thematic analysis produced a taxonomy with four main types of problematic machine behavior: discrimination, distortion, exploitation, and misjudgement. We propose this taxonomy as a potential high-level framework for the different harms caused by algorithmic systems, \added{although future audit work will likely diagnose additional behaviors that could expand this taxonomy}.

Discrimination is an important central focus of algorithm audit studies, and 21 studies in our review specifically audited for discrimination. As noted by Sandvig et al. \cite{Sandvig2014}, while "the word `audit' may evoke financial accounting, the original audit studies were developed by government economists to detect racial discrimination in housing." Racial discrimination has continued to be a key focus of algorithm audits, and of critical algorithm studies in general, as scholars have introduced different terms for racially discriminatory algorithms: "coded gaze" (by \citet{Buolamwini2018}), "digital poorhouse" (by \citet{Eubanks2018Inequality}), "algorithms of oppression" (by \citet{Noble2018}), "algorithmic inequity" (by \citet{Wachter-Boettcher2017}), and "the New Jim Code" (by \citet{Benjamin2019Race}), among others. In addition to racial discrimination, discrimination can also occur on the basis of age, sex, gender, location, socioeconomic status, and/or intersectional identities, all of which have been the subject of algorithm audits, as discussed in Section \ref{sec-discrimination}.

While discrimination was the central focus of algorithm audits reviewed, overall, more studies (N=29) in our review focused on distortion (Section \ref{sec-distortion}), especially for search algorithms (N=18) and recommender systems (N=7). Broadly, these audit studies scrutinize curated media from algorithmic systems, checking for distortion in terms of political partisanship, false information, suppression of sources, and more. The third behavior was exploitation. While many scholars have expressed concern over exploitation on algorithmic systems, especially in terms of personal information (e.g. see \citet{Zuboff2015}, \citet{Couldry2019}, and \citet{West2019}), only five studies addressed how algorithmic systems exploit media content and/or personal information (Section \ref{sec-exploitation}). Finally, the more general "misjudgement" behavior was the focus of seven studies in our review (Section \ref{sec-misjudgement}), all of which audited algorithmic systems for errant decisions or classifications, without exploring more specific harms related to discrimination, distortion, or exploitation. The following subsections discuss all four behaviors in more detail.

\begin{center}
\begin{table}[]
\small{

\begin{tabularx}{0.8\textwidth}{l|c|c|c|c|l}
Year                 & \textbf{Discrimination} & \textbf{Distortion} & \textbf{Exploitation} & \textbf{Misjudgement} & \textbf{Total} \\
\hline
2012                  & 1              &            &              &              & 1           \\
2013                  & 3              & 1          &              &              & 4           \\
2014                  & 1              &            &              &              & 1           \\
2015                  & 1              & 1          & 1            &              & 3           \\
2016                  & 1              & 1          &              &              & 2           \\
2017                  & 2              & 3          & 2            & 1            & 8           \\
2018                  & 3              & 9          & 1            & 1            & 14          \\
2019                  & 7              & 10         & 1            & 4            & 22          \\
2020*                  & 2*              & 4*          &              & 1*            & 7*           \\
\hline
Total                 & 21             & 29         & 5            & 7            & 62         
\end{tabularx}
}
\caption{Number of studies, by year and behavior.}
\label{table-year-behavior}
\end{table}
\end{center}

\begin{center}
\begin{table}[]
\small{

\begin{tabularx}{0.97\textwidth}{l|c|c|c|c|l}
Domain                 & \textbf{Discrimination} & \textbf{Distortion} & \textbf{Exploitation} & \textbf{Misjudgement} & \textbf{Total} \\
\hline
Search              & 5              & 18         & 2            &              & 25          \\
Advertising         & 3              & 2          & 3            & 4            & 12          \\
Recommendation      & 1              & 7          &              &              & 8           \\
Pricing             & 5              &            &              &              & 5           \\
Vision              & 5              &            &              &              & 5           \\
Criminal Justice    & 1              &            &              & 3            & 4           \\
Language Processing & 1              & 1          &              &              & 2           \\
Mapping             &                & 1          &              &              & 1           \\

\hline
 Total         & 21             & 29         & 5            & 7            & 62  
\end{tabularx}
}
\caption{Number of studies, by domain and behavior.}
\label{table-domain-behavior}
\end{table}
\end{center}

\subsection{Discrimination}\label{sec-discrimination}

This review defined \textit{discrimination} as an algorithmic system disparately treating or disparately impacting people with respect to their race, age, sex, gender, location, socioeconomic status, and/or intersectional identity. This section details how discrimination manifested in advertising algorithms, computer vision algorithms, search algorithms, and pricing algorithms. \added{Importantly, algorithmic discrimination can happen through allocation or through representation \cite{Barocas2017}.

Discrimination is often thought of in terms of allocation, and especially for "a narrow set of goods, namely rights, opportunities, and resources" \cite{Hoffmann2019}. However, discrimination can also play out through representational disparities, as evidenced in the work by \citet{Noble2013} and \citet{Kay} and further discussed in \ref{subsec-search}. Representational discrimination is sometimes referred to as "bias," however, that word has vastly different connotations and definitions in statistics (e.g. estimator bias), machine learning (e.g. bias term), psychology (e.g. cognitive bias), physics (e.g. tape bias), and other fields related to computing. This review therefore refrains from using the word "bias" whenever possible, recognizing the fact that public-facing algorithms exist in real-world social contexts where bias leads to harmful discrimination.}

\begin{table}
\rowcolors{1}{white}{gray!15}
\small{
\begin{tabularx}{\textwidth}{l c X X X X}
\toprule
              Reference &  Year &                                    Method &               Domain &	Class	&                                                           Organization \\
\midrule
    \citet{Mikians2012} &  2012 &                   \makecell{Sock puppets} &              Pricing &   \makecell{Class\\Geography} &                                          \makecell{Google \\ Bing} \\
   \citet{Mikians2013a} &  2013 &                  \makecell{Crowdsourcing} &              Pricing &   Geography &                   \makecell{Amazon \\ Hotels.com \\ Other websites} \\
    \citet{Sweeney2013} &  2013 &                  \makecell{Direct scrape} &          Advertising &   Race &                                                   \makecell{Google} \\
      \citet{Noble2013} &  2013 &                  \makecell{Direct scrape} &               Search &   Race &                                                   \makecell{Google} \\
         \citet{Hannak} &  2014 &  \makecell{Crowdsourcing \\ Sock puppets} &              Pricing &   \makecell{Generic\\Device} &                             \makecell{WalMart \\ Expedia \\ Orbitz} \\
            \citet{Kay} &  2015 &                  \makecell{Direct scrape} &               Search &   Gender &                                                   \makecell{Google} \\
       \citet{Chen2016} &  2016 &                  \makecell{Direct scrape} &              Pricing &   Generic &                                                   \makecell{Amazon} \\
  \citet{Eriksson2017a} &  2017 &                   \makecell{Sock puppets} &       Recommendation &   Gender &                                                  \makecell{Spotify} \\
        \citet{Hannaka} &  2017 &                  \makecell{Direct scrape} &               Search &   Intersectional &                                     \makecell{TaskRabbit \\ Fiverr} \\
  \citet{Hupperich2018} &  2018 &                   \makecell{Sock puppets} &              Pricing &   Geography &              \makecell{Booking.com \\ Hotels.com \\ Other websites} \\
          \citet{Chena} &  2018 &                  \makecell{Direct scrape} &               Search &   Gender &                       \makecell{Indeed \\ Monster \\ CareerBuilder} \\
 \citet{Buolamwini2018} &  2018 &                  \makecell{Direct scrape} &               Vision &   Intersectional &                               \makecell{Microsoft \\ IBM \\ Face++} \\
           \citet{Raji} &  2019 &                 Carrier puppet &               Vision &   Intersectional &                                         \makecell{Amazon\\Kairos} \\
    \citet{Barlas2019a} &  2019 &                  \makecell{Direct scrape} &               Vision &   \makecell{Gender\\Race} &                                          \makecell{(Same as\\below)} \\
   \citet{Kyriakou2019} &  2019 &                  \makecell{Direct scrape} &               Vision &   \makecell{Gender\\Race} & \makecell{(Same as\\below)} \\
    \citet{DeVries2019} &  2019 &                  \makecell{Direct scrape} &               Vision &   Geography &          \makecell{Amazon \\ Google \\ IBM \\ Microsoft \\ Clarifai} \\
 \citet{Lambrecht2019a} &  2019 &                 \makecell{Carrier puppet} &          Advertising &   Gender &        \makecell{Facebook \\ Google \\ Instagram \\ Twitter} \\
   \citet{CemGeyik2019} &  2019 &                           \makecell{Code} &               Search &   Gender &                                                 \makecell{LinkedIn} \\
          \citet{Tolan} &  2019 &                           \makecell{Code} &     Criminal Justice &   \makecell{Gender\\Race} &                                         \makecell{Catalonia,\\Spain} \\
    \citet{Asplund2020} &  2020 &                   \makecell{Sock puppets} &          Advertising &   \makecell{Gender\\Race} &                                                   \makecell{Google} \\
        \citet{Sap2020} &  2020 &                  \makecell{Direct scrape} &  \makecell{Language \\Processing} &  Race &                                    \makecell{Jigsaw\\(Google)} \\
\bottomrule
\end{tabularx}

}
\caption{Papers included in the review that audited for discrimination. Sorted by year.}
\label{results-table-discrimination}
\end{table}

\subsubsection{Discrimination in Advertising}
Researchers have mainly found discrimination in advertising with regards to \textit{who} is searching. \citet{Datta2015a} showed that protected classes directly affect ad targeting on Google, and later used this evidence to argue that platforms' immunity to Title VII should be reconsidered \cite{Datta2016}. Corroborating these findings, \citet{Cabanas2018} showed that some targeting features explicitly rely on protected and/or sensitive personal attributes such as sexual orientation, political preferences, and religious beliefs. Most recently, \citet{Asplund2020} conducted a sock-puppet audit which yielded evidence of "differential treatment in the number and type of housing ads served based on the user’s race, as well as bias in property recommendations based on the user’s gender." For example, sock puppets emulating Caucasians saw more housing-related advertisements, while sock puppets emulating African Americans saw more ads for predatory rent-to-own programs \cite{Asplund2020}. Algorithmic advertising has also been shown to discriminate on the basis of sexual and/or gender identity. \citet{Lambrecht2019a} found that Facebook advertisements for STEM career ads, even with gender-neutral targeting and bidding options, reached fewer women than men.

An earlier audit in our review showed an additional way that advertising can discriminate \added{through representational harms}, regardless of who is searching. The study, conducted by \citet{Sweeney2013}, showed through case studies and a large-scale analysis of 2,184 name searches that "ads suggesting arrest tend to appear with names associated with blacks, and neutral ads or no ads appear with names associated with whites." We discuss this phenomenon further in the search section, in light of audits by \citet{Noble2013}. Overall, algorithmic advertising systems often enable harmful discrimination which poses significant risks to users' safety, privacy, and economic opportunity.

\subsubsection{Discrimination in Vision}
The audits reviewed show computer vision algorithms discriminating on the basis of sex, race, geography, and intersectional identities. Many audits in this area stem from the seminal "gender shades" study by \citet{Buolamwini2018}, which showed that facial analysis on darker-skinned females performed significantly worse than on light-skinned males. The study demonstrated that disparities often arise from imbalanced training data, for example, benchmark datasets exhibited "overrepresentation of lighter males, underrepresentation of darker females, and underrepresentation of darker individuals in general" \cite{Buolamwini2018}. Because computer vision training data often exhibits imbalance, and imbalanced data often creates discriminatory performance disparities, researchers have audited a variety of systems for these problematic disparities.

In the case of sexual discrimination, performance disparities occur when a system exhibits higher error rates for a given sexual identity. Similar to findings by \citet{Buolamwini2018}, \citet{Kyriakou2019} found that several commercial systems for image tagging perform worse on female faces. In terms of representation disparities, one of the earliest algorithm audits by \citet{Kay} found that women were consistently underrepresented in Google's image search results compared to metrics from the U.S. Bureau of Labor Statistics.

Racial discrimination in photographic technology has been present for decades in cinema and photography \added{(e.g. \cite{Latif2017}, see also chapter 3 in \textit{Race After Technology}, by \citet{Benjamin2019Race})}, and in some ways, this appears to continue with computer vision algorithms. In a well-publicized incident involving the Google Photos app, one headline summarized "Google Mistakenly Tags Black People as `Gorillas'" \cite{Barr2015}. While not always so overt and egregious, formal audits of computer vision systems have consistently surfaced similar discrimination. An audit of five commercial image tagging systems \cite{Kyriakou2019,Barlas2019a} found worse performance on people with dark skin, a finding consistent with the original gender shades study \cite{Buolamwini2018} as well as a follow-up study that found performance disparities in three other commercial systems \cite{Raji}.

More recently, some studies have found computer vision algorithms can discriminate on the basis of geography. As with sexual and racial discrimination, this may arise from skewed training data, in particular, many popular training datasets consist of mostly images from North America and Europe, with fewer images from the global south. This imbalanced training data leads to discriminatory performance disparities, as a recent study noted that "object-classification accuracy in recognizing household items is substantially higher for high-income households than it is for low-income households" \cite{DeVries2019}.

Finally, the original gender shades study by \citet{Buolamwini2018} demonstrated the importance of considering intersectional identities when identifying discrimination. The three commercial systems audited in the gender shades study exhibited \textit{some} disparities between male faces and female faces, as well as between darker-skinned and lighter-skinned faces, however, disparities amplified when they were considered simultaneously. For example, the Face++ error rate among lighter-skinned males was 0.8\%, while the error rate among darker-skinned females was 34.5\%. A follow-up study \cite{Raji} also used intersectional subgroups to audit for potential improvements in commercial vision algorithms, measuring corporations' success by their ability "to significantly reduce error gaps in the intersectional performance of their commercial APIs" \cite{Raji}. \citet{Crenshaw1989} coined the term intersectionality for exactly these kinds of disparities, and \citet{Ogbonnaya-Ogburu2020} recently discussed their importance to HCI research: "each person represents a unique and even potentially conflicting set of overlapping identities... we must be anti-essentialist and incorporate an understanding that these intersecting identities create unique contexts."

\subsubsection{Discrimination in Search}\label{subsec-search}
Within our review, one of the earliest and most important audits of discrimination in search algorithms was performed by \citet{Noble2013} in 2013. The audit vividly demonstrated how search engines can reinforce harmful stereotypes with respect to intersectional identities: in a Google search for "black girls," five of the top ten results were sexualized or pornified, while only three of the top ten results were "blogs focused on aspects of social or cultural life for Black women and girls." Noble developed these findings of representational discrimination in a book about search engines and racism, titled "Algorithms of Oppression" \cite{Noble2018}.

Another important work focused on \added{representational} discrimination is the aforementioned audit by \citet{Kay} focused on Google's image search. As mentioned in the context of discriminatory computer vision algorithms, the audit found that image search results exaggerated gender stereotypes when searching for occupations, such as "nurse" and "doctor," compared to baselines from the U.S. Bureau of Labor and Statistics. The authors suggest that this behavior "risks reinforcing or even increasing perceptions of actual gender segregation in careers," \cite{Kay} especially given that users focus on highly-ranked results \cite{Granka2004}.

While Google provides search results from across the internet, some search algorithms are specific to content within a given platform. Discrimination in these search algorithms can lead to \added{egregious allocative harms} on platforms that influence employment opportunities, and audit studies have directed their attention accordingly. \citet{Hannaka} investigated discrimination in search results on TaskRabbit and Fiverr, finding evidence for discrimination on the basis of gender and race. Namely, the audit found that "workers perceived to be Black tend to be shown at lower ranks relative to those perceived to be White" \cite{Hannaka}. Another study by \citet{Chena} audited Indeed, Monster, and CareerBuilder, also surfacing evidence of gender discrimination ("overall, men rank higher than women with equivalent features"). On a more positive note, \citet{CemGeyik2019} conducted a similar audit of discrimination in LinkedIn talent search, and demonstrated a method for improving gender representation on the platform based on notions of equal opportunity and demographic parity. In A/B testing, their fairness-aware search algorithm produced gender-representative search results for 95\% of all queries, compared to 50\% for the original search algorithm, and the fairness-aware search algorithm was then deployed to LinkedIn Recruiter users worldwide.

\subsubsection{Discrimination in Pricing}

Lastly, some studies have audited for discrimination in pricing algorithms. E-commerce websites such as Orbitz, Amazon, and Home Depot sometimes show different prices to different people for the same item -- a phenomenon known explicitly as \textit{price discrimination}. This behavior presents clear potential for harm, by disproportionately burdening some populations while systematically advantaging others. As \citet{Jiang2019a} point out, ethical pitfalls in price discrimination surface even in random A/B/N tests that do not explicitly discriminate on the basis of race or class, for example. And while price discrimination is often legal, laws in the United States prohibit similar kinds of opportunistic price changes \cite{Editors2020}, making algorithmic price discrimination an important behavior to capture and understand through algorithm audits. To this end, \citet{Mikians2012} conducted the earliest audit in our review, with similar follow-up studies conducted by \citet{Mikians2013}, \citet{Hannak}, \citet{Chen2016}, and \citet{Hupperich2018}.

These price discrimination audits were primarily concerned with \textit{detecting} price discrimination and demonstrating its extent, and they offer fewer details about the specific discriminatory behavior. For example, the audit by \citet{Chen2016} focused on proving the existence of dynamic pricing, noting that dynamic price discrimination threatens to disrupt the market, exacerbate inequalities between sellers, and create a confusing shopping experience for customers. In other words, price discrimination causes harms even if it is not definitively on the basis of race, age, sex, gender, location, socioeconomic status, and/or intersectional identity. Still, some audits provide more specific analyses of such discrimination, such as the study by \citet{Mikians2012} which found evidence of discrimination on the basis of location and class (e.g. "affluent customers" versus "budget conscious customers").

\subsection{Distortion}\label{sec-distortion}

The \textit{distortion} behavior refers to an algorithm presenting media in a way that distorts or obscures an underlying reality. In many ways, distortion is analogous to discrimination, but rather than discriminating against different groups of people, the algorithm discriminates against different groups of media. This may involve political partisanship, unpredictable fluctuation, dissemination of misinformation, and/or hyper-personalized content that can lead to "echo chambers." Distortion is mainly of interest with regards to search algorithms and recommendation algorithms.

\begin{table}
\rowcolors{1}{white}{gray!15}
\small{
\begin{tabularx}{\textwidth}{l c X X X}
\toprule
                 Reference &  Year &                                     Method &               Domain &                  Organization \\
\midrule
        \citet{Hannak2013} &  2013 &   \makecell{Crowdsourcing \\ Sock puppets} &               Search &             \makecell{Google} \\
 \citet{Kliman-Silver2015} &  2015 &                    \makecell{Sock puppets} &               Search &             \makecell{Google} \\
       \citet{Soeller2016} &  2016 &                    \makecell{Sock puppets} &              Mapping &             \makecell{Google} \\
       \citet{Eslami2017c} &  2017 &                   \makecell{Direct scrape} &               Search &        \makecell{Booking.com} \\
   \citet{Kulshrestha2017} &  2017 &                   \makecell{Direct scrape} &               Search &  \makecell{Twitter \\ Google} \\
      \citet{Snickars2017} &  2017 &                    \makecell{Sock puppets} &       Recommendation &            \makecell{Spotify} \\
    \citet{Robertson2018c} &  2018 &                   \makecell{Crowdsourcing} &               Search &             \makecell{Google} \\
       \citet{Andreou2018} &  2018 &                   \makecell{Crowdsourcing} &          Advertising &           \makecell{Facebook} \\
      \citet{Bechmann2018} &  2018 &                   \makecell{Crowdsourcing} &       Recommendation &           \makecell{Facebook} \\
     \citet{Puschmann2018} &  2018 &                   \makecell{Crowdsourcing} &               Search &             \makecell{Google} \\
  \citet{Chakraborty2018b} &  2018 &                    \makecell{Sock puppets} &       Recommendation &     \makecell{New York Times} \\
     \citet{Courtois2018b} &  2018 &                   \makecell{Crowdsourcing} &               Search &             \makecell{Google} \\
        \citet{Weber2018a} &  2018 &                            \makecell{Code} &       Recommendation &      \makecell{Not specified} \\
  \citet{Kulshrestha2019b} &  2018 &                   \makecell{Direct scrape} &               Search &  \makecell{Google \\ Twitter} \\
            \citet{Rieder} &  2018 &                   \makecell{Direct scrape} &               Search &            \makecell{YouTube} \\
         \citet{Lurie2019} &  2019 &                   \makecell{Direct scrape} &       Recommendation &             \makecell{Google} \\
           \citet{Trielli} &  2019 &                   \makecell{Direct scrape} &               Search &             \makecell{Google} \\
           \citet{Moe2019} &  2019 &                   \makecell{Direct scrape} &               Search &            \makecell{YouTube} \\
       \citet{Metaxa2019a} &  2019 &                   \makecell{Direct scrape} &               Search &             \makecell{Google} \\
         \citet{Jiang2019} &  2019 &                   \makecell{Direct scrape} &  \makecell{Language\\ Processing} &            \makecell{YouTube} \\
                \citet{Hu} &  2019 &                   \makecell{Direct scrape} &               Search &             \makecell{Google} \\
    \citet{Robertson2019a} &  2019 &                   \makecell{Direct scrape} &               Search &     \makecell{Google \\ Bing} \\
           \citet{Ali2019} &  2019 &                  \makecell{Carrier puppet} &          Advertising &           \makecell{Facebook} \\
     \citet{Cano-Oron2019} &  2019 &                   \makecell{Direct scrape} &               Search &             \makecell{Google} \\
          \citet{Lai2019a} &  2019 &                   \makecell{Crowdsourcing} &               Search &             \makecell{Google} \\
       \citet{Hussein2020} &  2020 &                    \makecell{Sock puppets} &               Search &            \makecell{YouTube} \\
       \citet{Ribeiro2020} &  2020 &                   \makecell{Direct scrape} &       Recommendation &            \makecell{YouTube} \\
            \citet{Bandya} &  2020 &  \makecell{Direct scrape \\ Crowdsourcing} &       Recommendation &              \makecell{Apple} \\
       \citet{Fischer2020} &  2020 &                   \makecell{Direct scrape} &               Search &             \makecell{Google} \\
\bottomrule
\end{tabularx}

}
\caption{Papers included in the review that audited for distortion. Sorted by year.}
\label{results-table-distortion}
\end{table}

\subsubsection{Distortion in Search}
Search audits were common among the papers reviewed, and many of them cited the "politics of search engines" \cite{Introna2000} and/or the "search engine manipulation effect" (SEME) \cite{Epstein2017a} as motivation. The SEME phenomenon involves a search engine shifting voting behavior in accordance with presented search results. For example, a large-scale experiment found that a search engine led to a 39\% increase in the number of subjects "who indicated that they would vote for the candidate who was favored by their search rankings" \cite{Epstein2017a}. SEME suggests that search engines exert some degree of influence over voting behavior and democratic outcomes, which appears to have guided search audits to focus on partisanship. After all, experimental evidence for the SEME prompts an urgent question: do real-world search engine results favor certain political candidates?

The audits reviewed occasionally surfaced evidence for small but statistically significant partisan leanings. One of the earliest studies showing such evidence was a set of case studies by \citet{Diakopoulos}, which found that Google searches for Republican candidates yielded a higher proportion of negative articles compared to searches for Democratic candidates, and overall the partisanship of search results leaned left. Several studies have replicated this small but significant left-leaning partisanship on Google, especially in terms of news items \cite{Trielli,Robertson2018c}. However, the result is open to many different interpretations. Also, some studies have found conflicting results. For example, while \citet{Robertson2018c} found news items exhibited a slight left-leaning partisanship, they also found that "Google's ranking algorithm shifted the average lean of [Search Engine Results Pages] slightly to the right." Furthermore, \citet{Metaxa2019a} observed "virtually no difference" in the distribution of source partisanship between queries for Democrats versus Republicans. Shifting the focus beyond just the "blue links," \citet{Hu} audited the "snippet" text that appears below the links on search result pages. They found that partisan cues in the snippet text tend to amplify the partisanship of the original web page, an effect that was consistent across query topics, left- and right-leaning queries, and different types of web pages (e.g. social media, news, sports).

Even in the absence of partisanship, search algorithms can distort media by limiting exposure diversity, which runs counter to the notion of "the Web as a public good," as argued by \citet{Introna2000}. This can occur through overall source concentration, and through hyper-personalization that may create "echo chambers." In terms of overall source concentration, an audit by \citet{Trielli} focused on the "Top Stories" carousel of Google's results page, finding a high concentration of sources (the top five sources of impressions per query were CNN, New York Times, The Verge, The Guardian, and Washington Post). The substantial dependence on mainstream media outlets has also been evidenced in studies of medical queries \cite{Cano-Oron2019}, political queries in Dutch-speaking regions \cite{Courtois2018b}, and political queries in the United States \cite{Kulshrestha2019b,Robertson2018c}. An audit by \citet{Lurie2019} provides an exception to this trend in source concentration, suggesting that Google "uses the 3rd position for exploration, providing users with unfamiliar sources," thus promoting lesser-known publishers. More recently, \citet{Fischer2020} found that national outlets still dominate Google search results overall, suggesting that the search algorithm "may be diverting web traffic and desperately needed advertising dollars away from local news" \cite{Fischer2020}.

Exposure diversity can also be limited through "echo chambers" that may arise when a search algorithm provides hyper-personalized results to users \cite{Bozdag2015}. The earliest personalization audit reviewed was by \citet{Hannak2013} in 2013, which found that on average only 11.7\% of results were personalized, although personalization occured more often in some query categories ("gadgets," "places," and "politics"). A later study by \citet{Kliman-Silver2015} showed that geolocation had the greatest effect on personalized results.

Generally, studies have shown limited evidence for the "echo chambers" that some scholars have feared. A study of 350 Dutch users found no substantial differences due to personalization \cite{Courtois2018b} across 27 socio-political queries. An audit by \citet{Robertson2018c} measured personalization on all different components of a search engine including the "people also ask for" box, "tweet cards," and "knowledge components." The full-page audit "found negligible or non-significant differences between the [Search Engine Results Pages]" of personalized (logged in to Google) and non-personalized (incognito) windows. These findings could be limited to socio-political queries -- as the earliest personalization audit showed, the degree of personalization varies across query categories \cite{Hannak2013}. Rather than queries for political candidates and topics, one study in New Zealand by \citet{Lai2019a} used queries that public officials "might perform in the course of their everyday work duties." Asking 30 participants to rate the relevance of personalized and non-personalized results, their findings suggest that up to 20\% of relevant results were removed due to personalization.

Lastly, in addition to partisanship and hyper-personalization, a search algorithm may be culpable of distortion if it disseminates low-quality media such as false information, junk news, or other opportunistic content. \citet{Hussein2020} explore this phenomenon on YouTube, finding that the search algorithm exhibited a "rabbit hole" effect: "people watching videos promoting misinformation are presented with more such videos in the search results." The effect was not present in search results for queries related to vaccines, which is notable given the high stakes of health information illustrated by \citet{DeChoudhury2014}.

\subsubsection{Distortion in Recommendation}
Distortion in recommendation is similar to distortion in search, and presents a similar level of public impact. Examples of public-facing recommender systems include news recommenders such as Google News, song recommenders such as Spotify radio, and product recommenders such as the "customers also bought" feature on Amazon. Audits of recommender algorithms have focused on hyper-personalization and echo chamber effects, as well as source concentration.

As with internet search engines, recommender systems have the potential to create echo chambers by recommending users a narrow, specific set of content. In a typical depiction of the phenomenon, a user reads recommended news articles with a specific viewpoint, forming a feedback loop with the recommender which continually reinforces that viewpoint. But as with search algorithms, the echo chamber phenomenon has failed to materialize in numerous studies examining Google News recommendations \cite{Haim2018,Nechushtai2019}, Facebook News Feed recommendations \cite{Bechmann2018,Bakshy2015}, and "Trending Stories" recommendations in Apple News \cite{Bandya}. One study of the New York Times website \cite{Chakraborty2018b} found evidence for a potential feedback loop on the "Recommendations for You" page. However, the study focused on a single, explicitly-personalized page of the website, and did not examine ideological or topical differences in the content. Lastly, personalization effects were also minimal in an audit of Spotify's radio recommendations -- not even liking all songs or disliking all songs had a significant impact on recommended songs \cite{Snickars2017}.

While audits have rarely surfaced significant echo chamber effects, they have found that recommendation algorithms often exhibit problematic source concentration. The recent study by \citet{Bandya} found that the top three sources of "Trending Stories" in Apple News accounted for 45\% of all recommendations. Similar source concentration occurs on Google News. Echoing early hints from \citet{Haim2018}, \citet{Nechushtai2019} found that 35\% of all recommended Google News stories came from just three sources. As noted by \citet{Introna2000} "systematically giving prominence to some at the expense of others" \added{in these contexts} is often inherently inequitable and problematic.

\subsection{Exploitation}\label{sec-exploitation}

This study considered \textit{exploitation} to be the inappropriate use of content from other sources and/or sensitive personal information from people. While discrimination applies to people and distortion applies to media, algorithms can exploit either people or media. For example, an algorithm implicated in exploitation may infer sensitive personal information from users without proper consent, or feature content from an outside source without attribution. \added{This review found five studies that focused on exploitation in algorithmic systems.}

\subsubsection{Exploitation in search}
Search engines face heightened scrutiny and audit attention for their reliance on -- and potential exploitation of -- outside content. Indeed, almost all studies reviewed in this area present evidence that search engines depend heavily upon user-generated content (UGC) and journalism content, which presents a number of potential problems entwined with copyright laws and monopoly power \cite{Dreyfuss2019}. 

Notably, audits show that user-generated content from websites like Wikipedia, Twitter, and StackOverflow improve click-through rates on Google \cite{McMahon2017}. Considering that UGC adds value to Google search in the form of improved click-through rates, \citet{Vincent2019} conducted an audit to quantify the extent of Google's UGC reliance, finding that Wikipedia content, created through voluntary labor, appears in 81\% of results pages for queries trending in Google Trends, and 90\% of results pages for queries about controversial topics (e.g. "Death Penalty," "Under God in the Pledge," "Israeli-Palestinian Conflict"). The study also found that the "News Carousel" appeared in the majority popular and trending queries, often in the top three Google results. As noted in the distortion section (\ref{sec-distortion}), Google's reliance on news outlets is highly concentrated \cite{Trielli,Kulshrestha2019b,Robertson2018c}, and their results often exclude local news \cite{Fischer2020}.

\subsubsection{Exploitation in advertising}
Advertising algorithms use sophisticated experimentation and optimization techniques to deliver personalized advertisements, in some cases leading them to exploit sensitive personal information. Distinct from algorithm audits for discrimination, audits for exploitation focus on the appropriation of sensitive personal information, especially without users' consent. A clear example of this was surfaced in the audit by \citet{Datta2015a}: agents with browsing history related to substance abuse were shown a different distribution of ads, but this trait was not shown in Google's Ad Settings. In other words, beyond discriminatory targeting (male agents were more likely than female agents to see ads for high paying jobs), Google's advertising algorithms also tracked and exploited private, sensitive browsing behavior to target users with ads.

With an eye toward the General Data Protection Regulation in the European Union, \citet{Cabanas2018} noted that Facebook "should obtain explicit permission to process and exploit sensitive personal data" for commercial gain. Yet their audit showed that Facebook was exploiting sensitive information in their advertising algorithms, allowing advertisers to target people in categories such as "interested in homosexuality," "interested in Judaism," "interested in tobacco," and more. Again, these categories will likely lead to discrimination and other harms, \added{though these studies highlight the problematic behavior of} non-consensually inferring and exploiting these sensitive attributes for targeted advertising, as warned about by \citet{Zuboff2015}, \citet{Couldry2019}, \citet{West2019}, and other scholars.

\begin{table}
\rowcolors{1}{white}{gray!15}
\small{
\begin{tabularx}{\textwidth}{l c X X X}
\toprule
           Reference &  Year &                    Method &       Domain &         Organization \\
\midrule
  \citet{Datta2015a} &  2015 &   \makecell{Sock puppets} &  Advertising &    \makecell{Google} \\
  \citet{Mahler2017} &  2017 &  \makecell{Direct scrape} &  Advertising &   \makecell{Spotify} \\
 \citet{McMahon2017} &  2017 &  \makecell{Crowdsourcing} &       Search &    \makecell{Google} \\
 \citet{Cabanas2018} &  2018 &  \makecell{Crowdsourcing} &  Advertising &  \makecell{Facebook} \\
 \citet{Vincent2019} &  2019 &  \makecell{Direct scrape} &       Search &    \makecell{Google} \\
\bottomrule
\end{tabularx}

}
\caption{Papers included in the review that audited for exploitation. Sorted by year.}
\label{results-table-exploitation}
\end{table}

\subsection{Misjudgement}\label{sec-misjudgement}
Misjudgement was defined generally as an algorithm making incorrect predictions or classifications. As noted earlier, misjudgement can often lead to discrimination, distortion, and/or exploitation, but some studies in our review focused on this initial error of misjudgement without exploring second-order problematic effects. \added{Seven studies in the review focused on misjudgement, and were clustered in two areas: criminal justice and advertising.}

\begin{table}
\rowcolors{1}{white}{gray!15}
\small{
\begin{tabularx}{\textwidth}{l c X X X}
\toprule
              Reference &  Year &                    Method &            Domain &                                                     Organization \\
\midrule
       \citet{Duwe2017} &  2017 &  \makecell{Direct scrape} &  Criminal Justice &                                             \makecell{Minnesota} \\
  \citet{Tschantz2018a} &  2018 &  \makecell{Crowdsourcing} &       Advertising &                                                \makecell{Google} \\
 \citet{Venkatadri2019} &  2019 &  \makecell{Crowdsourcing} &       Advertising &  \makecell{Facebook\\ Acxiom\\ Epsiolon\\ Experian\\ Oracle (Datalogix)} \\
       \citet{Duwe2019} &  2019 &  \makecell{Direct scrape} &  Criminal Justice &                                             \makecell{Minnesota} \\
     \citet{Bashir2019} &  2019 &  \makecell{Crowdsourcing} &       Advertising &                        \makecell{Google\\ Facebook\\ Oracle\\ Neilsen} \\
   \citet{Matthews2019} &  2019 &           \makecell{Code} &  Criminal Justice &                                         \makecell{New York City} \\
      \citet{Silva2020} &  2020 &  \makecell{Crowdsourcing} &       Advertising &                                              \makecell{Facebook} \\
\bottomrule
\end{tabularx}

}
\caption{Papers included in the review that audited for misjudgement. Sorted by year.}
\label{results-table-misjudgement}
\end{table}

\subsubsection{Misjudgement in Criminal Justice}
The use of algorithms in criminal justice presents alarming potential for harm. After ProPublica's 2016 "Machine Bias" report \cite{Kirchner2016} suggested that a publicly-used recidivism prediction instrument (RPI) discriminated on the basis of race, researchers discussed and explored appropriate fairness criteria and methods for balancing error rates \cite{Chouldechova2017,Flores2016}. While these theoretical and methodological explorations are not included in our review, a handful of related papers met the inclusion criteria. Two studies \cite{Duwe2017,Duwe2019} focused on recidivism prediction systems in the state of Minnesota, and detailed methods to improving prediction accuracy. \citet{Tolan} provide a similar case study, showing that machine learning is more accurate but less fair (in terms of demographic parity and error rate balance) compared to statistical models. Also, in a code audit of a forensic DNA analysis system in New York City, \citet{Matthews2019} showed how an allegedly "minor" change to the algorithm resulted in substantial data-dropping that led to more inaccurate results: a small change to removal criteria ended up excluding more true contributors and included more non-contributors for the DNA samples analyzed, which led to misjudgments and inaccuracies.

As we discuss \added{later in section \ref{sec-risk-assessment}}, algorithms in the context of criminal justice do not need audits and incremental improvements as much as they need holistic reform \added{and abolition}, as suggested by \citet{Benjamin2019Race} and others. 

\subsubsection{Misjudgement in Advertising}
While advertising systems promise fine-grained targeting, audits show that targeted advertising algorithms often make misjudgements when inferring information about users, including demographic attributes and interest-based attributes. \citet{Tschantz2018a} found that among logged out users across the web, Google correctly inferred age for just 17\% of females and 6\% of males. Furthermore, a 2019 study by \citet{Venkatadri2019} showed a substantial amount of third-party advertisers' data about users is "not at all accurate" (40\% of attributes among the 183 users surveyed) -- some users labeled as corporate executives were, in fact, unemployed. These inaccuracies also apply to interest-based attributes, as \citet{Bashir2019} show in their audit of Facebook, Google, Oracle, and Neilsen. Asking users to review their "interest profiles" on these platforms, participants only rated 27\% of the interests in their profiles as strongly relevant. Taken in concert, these audits suggest that advertisers who purchase targeted advertisements may often fail to reach their intended audience on platforms like Facebook and Google. While advertisers would find these misjudgements undesirable, notably, the same behavior may be desirable to users, who find it invasive when algorithms infer sensitive attributes without their consent \cite{Zhang2014,Wang2012}.

\section{Results for RQ2: Future Algorithm Audits}
While the previous section addressed how algorithm audits have \added{diagnosed} problematic machine behavior in some areas, this section points out areas that require further research attention. These areas were selected based on \added{public impact as well as} relative coverage reflected in Table \ref{table-domain-behavior}, which tabulates studies by domain and problematic behavior. Importantly, this means that research areas in this section have received \textit{less} attention, rather than \textit{no} attention. This section thus includes studies that address or begin to address the topics at hand, as a starting point for future audits. As with any literature review, these findings are subject to potential coverage biases. Some studies may have been errantly excluded due to the keyword search, mistakes in the coding process, and/or inaccessible publication.

\subsection{Remaining Work: Discrimination}
The audits reviewed in this paper shed significant light on the path forward for studying discrimination in algorithmic systems. Previous audits provide a number of helpful precedents for future research, and hint at potential issues in \added{under-explored areas}, which deserve further research attention.

The areas reported in Section \ref{sec-discrimination} (discriminatory advertising, discriminatory pricing, discriminatory search, and discriminatory vision) received substantial audit attention, while still pointing to a need for further audits. As the economic engine behind large technology platforms, advertising in particular deserves further scrutiny, especially as companies like Facebook espouse changes intended to mitigate discriminatory harms (e.g. by removing the "multicultural affinity" targeting option in August 2020 \cite{Hutchinson2020}). For example, future algorithm audits might ask: do Facebook's measures mitigate discrimination, or obfuscate it? Even without overtly discriminatory targeting options, algorithmic targeting options that rely on income, location, and/or "lookalike audiences" may insidiously reproduce discriminatory behavior.

Similarly, price discrimination algorithms deserve further audit attention beyond merely \textit{detecting} dynamic pricing. In their piece describing the study of "machine behavior," \citet{Rahwan} explicitly suggest price discrimination as a research area. Following through on early evidence that pricing algorithms discriminate on the basis of location and class \cite{Mikians2012}, future audits should seek evidence for more specific discrimination on the basis of race, age, sex, gender, and/or intersectional identities. 

Accounting for intersectional identities is particularly important for future audits, as others such as \citet{Hoffmann2019} have pointed out, especially to avoid "fairness gerrymandering" \cite{Kearns2018}. In the words of \citet{Ogbonnaya-Ogburu2020}, HCI research aimed at reducing inequality (including algorithm audits) "must be anti-essentialist and incorporate an understanding that these intersecting identities create unique contexts." Ignoring intersectional identities means painting with too broad a brush, often aggregating and thus obfuscating the harmful disparities that affect people in their real-world, multifaceted identities. \added{In other words, audit studies must center intersectionality in order to work toward algorithmic systems that "explicitly dismantle structural inequalities" \cite{Eubanks2018Inequality}.}

This review also suggests two primary \textit{domains} which have received negligible audit attention with respect to discrimination: language processing algorithms and recommendation algorithms. Just one study in the review (by \citet{Sap2020}) audited a public-facing language processing algorithm, finding that the Perspective API tool from Jigsaw/Alphabet \added{exhibited} racial discrimination. A number of related studies find racial discrimination in \textit{non-public} language algorithms, which were thus not included in the review. Future language processing audits may benefit from taking a similar approach to computer vision audits, targeting corporate APIs rather than generic algorithms. For example, amidst the COVID-19 pandemic, some platforms increased their reliance on automated moderation, leading to inconsistent policies and frustration for many users \cite{Matsakis}. As demonstrated by \citet{Raji}, "publicly naming and disclosing performance results of biased AI systems" can directly improve real-world systems \added{and benefit the public}, in addition to serving the academic community. 

Recommendation algorithms were the second domain that received relatively little audit attention with respect to discrimination, and they would likely benefit from following precedents in other domains. One exception was a study of discriminatory "gendered streams" on Spotify, in which \citet{Eriksson2017a} found that approximately 8 out of 10 recommended artists from Spotify were male (the authors identified gender presentation for each artist based on pronouns, names, and images). This kind of discrimination may exist in a variety of recommendation algorithms, such as social media algorithms that recommend who to follow, apps that recommend restaurants \cite{Li2020}, business reviews \cite{Eslami2017c}, and more. Algorithm audits targeting recommender systems will benefit from the rich literature in search discrimination, on platforms like LinkedIn, Monster, and CareerBuilder (e.g., \cite{CemGeyik2019,Chena}). They will also become increasingly important as recommender systems become more common and influential in the \added{public} digital ecosystem.

\subsubsection{Discriminatory Risk Assessment}\label{sec-risk-assessment}

Algorithmic systems used in criminal justice have been scrutinized since ProPublica's 2016 report entitled "Machine Bias" \cite{Kirchner2016}, which suggested that a publicly-used recidivism prediction instrument (RPI) exhibited racial discrimination. The report prompted many discussions and explorations of fairness criteria and methods for balancing error rates (e.g., \cite{Chouldechova2017,Flores2016}), but surprisingly, this literature review found just one study that audited risk assessment algorithms for discrimination, authored by \citet{Tolan}. The audit found that machine learning models "tend to discriminate against male defendants, foreigners, or people of specific national groups." For example, the study showed that machine learning models were twice as likely to incorrectly classify people who were not from Spain (the dataset came from the Spanish community of Catalonia). Notably, three other studies in the review (\cite{Matthews2019,Duwe2017,Duwe2019}) audited algorithmic systems used in criminal justice, but focused on general misjudgement rather than discrimination.

In the context of criminal justice, researchers might benefit from an abolitionist rather than a reformist approach, as articulated by \citet{Benjamin2019Race} and others. Auditing for statistical disparities may be important in some cases, but often, the very use of algorithms in this context is at odds with any notion of justice or fairness. This is especially true in the United States, given the historic racial discrimination in many facets of the criminal justice system \cite{Alexander2010}. For example, if an algorithm is being used to inform an already oppressive process, the focus should be on abolishing that algorithm rather than auditing and improving it. \citet{Keyes2019a} poignantly illustrate this point through a satirical proposal for making a "fair, accountable, and transparent" algorithm to determine "which elderly people are rendered down into a fine nutrient slurry."

\subsection{Remaining Work: Distortion}
Audits reviewed in this study provided important insights about the ways that search and recommendation algorithms can distort digital media. Many studies focus on partisanship, echo chamber effects, and source concentration. These audits should continue, as they provide key insights about how algorithms exercise power in the media ecosystem. Future audits may also benefit from exploring other use cases, systems, and domains.

This review corroborates the suggestion by \citet{Mustafaraj2020} that existing audit literature has only accounted for a narrow set of use cases. Future audits should strive for user-centered audits, and may benefit from scoping to specific use cases. For example, such audits can build on the work of \citet{Lai2019a} which scoped to public officials' use of search algorithms, and the work of \citet{Mustafaraj2020} which outlines methods for \textit{voter}-centric audits. Also, \citet{Fischer2020} scoped their audit to local news, a topic which deserves more audit attention given the current crisis in local journalism and its threat to impacted communities (e.g. \cite{Abernathy2018,Darr}).

The algorithm audit literature should also expand their efforts to different systems \added{that may exhibit distortion}. Again, it is important to conduct ongoing audits of popular systems and platforms such as Google search, but researchers must also be vigilant in directing their attention to those that are new and upcoming. For example, the Google Discover recommendation algorithm has driven significant web traffic since its introduction in 2018 \citet{Willens2019}, but no audits in this review examined the system. Similarly, TikTok's rapid growth around the world \cite{Sun2020} presents new high-impact algorithms that call for audit attention. Lastly, personalization mechanisms in digital maps present a number of problematic potential distortions, such as the misrepresentation of international borders explored by \citet{Soeller2016}.

Distortion should also be audited in different domains, such as advertising and language. While echo chambers have been studied in the context of search and recommendation algorithms, advertising algorithms were the subject of just one echo chamber audit in this review. In the study, \citet{Ali2019} found evidence that Facebook's advertising algorithms "preferentially exposes users to political advertising." In some cases, even when advertisers targeted users with opposing viewpoints, Facebook preferred to show advertisements that aligned with the users' viewpoints. Given this initial evidence, as well as the pervasiveness of targeted advertising algorithms on today's platforms, future audits should further scrutinize these algorithms. In fact, more so than search and recommendation algorithms, advertising algorithms are \textit{premised} on hyper-personalized content that could create echo chambers, limiting the diversity of sources and content that users encounter. Advertising algorithms may also distort media by disseminating problematic content, such as false information or prohibited advertisements. As early work in this area, \citet{Silva2020} found that Facebook's advertising algorithm sometimes misjudged whether an ad was political, and showed how the system would violate political advertising laws in some jurisdictions.

Language algorithms present another important domain, with some early work signaling the need for more attention. \citet{Hu} showed that Google's algorithm for selecting search snippet text (which shows below the blue links) can distort the page it represents, for example, 54–58\% of snippets amplified partisanship, and 19–24\% of snippets exhibited inverted partisanship compared to the corresponding web page. With a similar focus on partisanship, an audit of comment moderation on YouTube by \citet{Jiang2019} dispelled a perception that YouTube's comment moderation practices were politically biased (rather, comments were more likely to be moderated if they were extremist, contained false content, or were posted after a video was fact-checked). But partisanship is not the only potentially distorting behavior, as some language systems can present misleading information. In an audit of Facebook's "why am I seeing this ad?" feature, \citet{Andreou2018} found that Facebook's linguistic explanations for targeted ads were often incomplete and sometimes misleading (similar to Google's lack of transparency demonstrated by \citet{Datta2015a}). Future audits should continue exploring these phenomena -- partisan language and misleading language -- in algorithms for various types of text summarization and language generation.

\subsection{Remaining Work: Exploitation}
Given the urgent concerns around exploitation on algorithmic systems, especially in terms of personal information, further audits of algorithmic exploitation \added{could provide important clarifications}. Considering the arguments and concerns articulated in concepts such as "surveillance capitalism" (by \citet{Zuboff2015}), "data colonialism" (by \citet{Couldry2019}), "data capitalism" (by \citet{West2019}), and others, researchers may benefit from focused audits that characterize specific types of exploitation and consequent harms inflicted by algorithmic systems. Given the methodological challenges associated with auditing exploitation, future audits in this area should look to successful early work by \citet{Datta2015a} and related projects in advertising (e.g., \cite{Cabanas2018, Venkatadri2019}).

One question in particular looms over this topic: how much money do companies make from exploitation? Some early work may help guide future audits, addressing the value of personal information and other types of content like news media.

In terms of personal information, \citet{GonzalezCabanas2017} explored how Facebook users perceived the economic value of their data. The study found that only 23\% of participants provided a close answer to the actual value of their personal data (about \$1 per month, estimated using Facebook's quarterly ad revenue and monthly active users in Q2 2016). Users' personal data also adds value to recommender systems. \citet{Vincent2019a} show that removing users' data "can bring recommender accuracy down to the levels of early recommender systems from 1999," thus, recommendation algorithms "are in fact a highly cooperative project between the public and companies." Even if the "data labor" used to power recommendation algorithms provides only a marginal increase in performance, it provides a massive increase in value. Netflix estimates a 2-4x increase in engagement, and about \$1 billion in revenue per year \cite{Gomez-Uribe2015}. Future algorithm audits should seek further clarification as to how platforms profit from exploiting users' personal information.

Similar clarifications are needed as to how platforms profit from news media and other content from journalism organizations. The News Media Alliance addressed this question in 2019, but extrapolated revenue share from a 2008 statistic, making the estimate of Google's revenue from news publishers (\$4.7 billion) questionable. Also, Google announced in 2020 it will pay publishers more than \$1 billion over the next three years to license news content. Moving forward, the ongoing debates about platforms and news media may benefit from audit studies that analyze the distribution of these funds, how equitable the distributions are, and how publishers are impacted -- especially struggling local publishers which may be particularly disadvantaged by Google \cite{Fischer2020}.

\subsection{Remaining Work: Misjudgement}
Future audits exploring misjudgement should generally opt to address more specific harms. Errant algorithms are often problematic merely by virtue of being errant, however, as detailed in the sections about discrimination, distortion, and exploitation, there is often a more specific problematic behavior associated with a simple error or misjudgement.

For example, a number of studies in our review addressed misjudgements in targeted advertising platforms, suggesting that advertising algorithms hold inaccurate targeting information about many users. In future audits, researchers may explore how these misjudgements specifically harm users and advertisers. Users may be unsettled when encountering a distorted characterization of their interests in the form of inaccurately targeted advertisements (as \added{evidenced} by \citet{Wang2012}) -- in fact, a Pew Research Study \cite{Center} found that "about half of Facebook users say they are not comfortable when they see how the platform categorizes them." Also, advertisers may be alarmed to know they spent money on inaccurately targeted advertisements, for example, aiming to reach corporate executives but in actuality reaching people experiencing unemployment (an error surfaced twice by \citet{Venkatadri2019}). In addition to simply identifying the initial error, future audits should focus on these potential harms to users, advertisers, \added{and other potential stakeholders.}

\subsection{Methods and Organizations for Future Audits}
\subsubsection{Methods}
While this review is primarily organized by behavior and domain, it is also important to note some promising areas for future work based on methods used and organizations audited. In terms of methods, Table \ref{table-method-organizations}a shows that direct scraping was the most-used audit method (N=30) in the papers we reviewed (N=62). As noted by \citet{Sandvig2014}, a key limitation with scraping audits is that there is no randomization or manipulation, making them a primarily useful for descriptive tasks. The audits that used crowdsourcing in our review (N=16) did not suffer from this limitation, since crowdsourced audits provide real-world data and can even allow for some causal inferences. While the carrier puppet method was used less often (N=3), this may be desirable given that such audits may disrupt real-world systems in the process of auditing.

Code auditing appears to be under-explored. This may be due to the fact that many algorithms of interest to researchers are proprietary and opaque, with no code to audit in the first place. However, code audits may be useful for some high-impact and open-source systems, such as applications by Wikipedia and DuckDuckGo.\footnote{\url{https://github.com/wikimedia}, \url{https://github.com/duckduckgo}} Generally, however, researchers can expect to continue using other methods when auditing systems from Facebook, Google, Amazon, and other corporations, since these systems are often proprietary and are unlikely to be open-sourced.

\setlength{\tabcolsep}{10pt}
\begin{table}[]
\centering
\small{
\begin{tabular}{lr}
\toprule
                                    Method &  Papers Reviewed \\
\midrule
                  Direct scrape &               30 \\
                  Crowdsourcing &               16 \\
                   Sock puppets &               12 \\
                    Code audit &                4 \\
                 Carrier puppet &                3 \\
\bottomrule
\\
\\
\\
\\
\\
\end{tabular}

}
\small{
\begin{tabular}{lr}
\toprule
                                    Organization &  Papers Reviewed \\
\midrule
Google           & 30 \\
Facebook         & 8  \\
Amazon           & 5  \\
YouTube          & 5  \\
State/Government & 4  \\
Twitter          & 3  \\
Spotify          & 3  \\
LinkedIn         & 1  \\
Instagram        & 1  \\
Apple            & 1  \\
\bottomrule
\end{tabular}

}
\\
\caption{Studies reviewed, tabulated by method (left) and organization (right).}
\label{table-method-organizations}
\end{table}

\subsubsection{Organizations}
Based on our review, the organizations audited have received skewed attention. Google  has been the subject of the most audits (N=30), followed by Facebook (N=8), Amazon (N=5), and YouTube (N=5), as shown in Table \ref{table-method-organizations}. The focus on Google is warranted given the influence of its search algorithms. At the same time, there also appears to be a gap in audit attention toward other influential organizations, especially social media platforms such as Facebook, Twitter, Instagram, and LinkedIn. This may be due in part to the legal challenges with scraping these platforms, noted by \citet{Sandvig2014}. With these challenges in mind, future audits may benefit from leveraging real-world data from end users, namely through crowdsourcing. Audits may also benefit from new data sources, such as Twitter's new academic access\footnote{\url{https://developer.twitter.com/en/solutions/academic-research}} and Facebook's dataset of all URLs shared more than 100 times \cite{Messing2020}.

\subsubsection{Replication}
Replication also presents an important area for future algorithm audits. While the findings in previous audit work are important and compelling, they may become quickly outdated given the frequent updates to algorithmic systems. Results may also change if researchers use different methods (e.g. crowdsourcing instead of sock puppets) or explore algorithms from different organizations (e.g. Bing's search algorithm instead of Google's search algorithm). These efforts would coincide with a broader push in computing research for replication studies \cite{Chi2011}, demonstrated clearly in the "repliCHI" workshop \cite{Wilson2013}.

Fortunately, researchers have already demonstrated a commitment to open science practices that will help enable replication audits. To name a few examples, \citet{Buolamwini2018} released a facial analysis dataset balanced by gender and skin type, \citet{Robertson2019} released scraping tools for recursive algorithm interrogation (RAI) and scraping web searches, and \citet{Barlas2019a} published their "SocialB(eye)as" dataset for exploring biases in computer vision algorithms.

\subsection{Important Ingredients for Future Audits}
To help guide future work in algorithm auditing, here we note the common ingredients for a successful audit that clearly demonstrates how problematic machine behavior affects the public. \added{While different systems, different contexts, and different research questions prompt different methods}, we identify at least four important common ingredients for a successful algorithm audit: a public-facing algorithmic system, suspected problematic behavior(s), a compelling baseline, and metric(s) to quantify the problematic behavior(s).

\subsubsection{Public-Facing Algorithmic System}
Given the growing number of opaque algorithmic systems that are already \added{deployed to} the public, audits that focus on these public-facing systems are especially important. \added{In the absence of "an FDA for algorithms" \cite{Tutt2017}, researchers can continue conducting audits to clarify the safety of public-facing algorithms, and help determine if and how they serve the public good.} In some cases, public-facing does not mean the system is publicly-deployed, only that it holds imminent or potential harms to the public.

\subsubsection{Suspected Problematic Behavior(s)}
With a public-facing \added{algorithmic} system identified, an audit study should define suspected problematic behavior(s), asking "how [social problems] manifest in technical systems" \cite{Abebe2019} in the most specific terms possible. Exploratory analysis can be helpful in some cases, but algorithmic systems tend to lend themselves to specific behavioral patterns, which should be the focus of audit studies. 

\subsubsection{A Compelling Baseline}
Audits should establish compelling baselines that convincingly answer the question: "compared to what?" For example, one of the earliest audits reviewed compared occupational gender representation compared to statistics from the U.S. Census Bureau \cite{Kay}. In other cases, such as the LinkedIn Talent Search audit by \citet{CemGeyik2019} used \textit{equality of opportunity} as a guiding theoretical framework to define a compelling baseline. Thus, compelling baselines need not come from the real world, and in many cases they should come from theoretical frameworks such as decolonialism, reparatory justice, feminist ethics of care, and justice as fairness, among others. For example, \citet{Costanza-Chock2018} explores some of these frameworks through the lens of \textit{Design Justice}, \citet{d2020datafeminism} outline a theory of \textit{Data Feminism}, and \citet{Mohamed2020} introduce a framework for "Decolonial AI." All such frameworks and concepts may be useful to establishing compelling baselines \added{beyond parity, group fairness, and other statistical metrics}.

\subsubsection{Metric(s) to Quantify Problematic Behavior(s)}
Finally, audits should present clear metrics that quantify the problematic behavior. Unfortunately, auditing presents the potential for a kind of "p-hacking" that has plagued other scientific disciplines \cite{Head2015,Hesselmann2017}. Algorithm auditors can almost always find some metric that suggests inequity, discrimination, or other problematic behavior. (On the flipside, organizations can \added{perform "fairness gerrymandering" \cite{Kearns2018}} and find some metric that suggests equity, equal treatment, etc. \cite{Flores2016}) Thus, audit studies can benefit from compelling baselines and robust, transparent, meaningful metrics to quantify and compare against the baseline. These metrics should be crafted with great care, especially when accounting for "representational harms" that \added{can manifest more subtly than} "allocative harms" \cite{Barocas2017}.

\section{Discussion}\label{discussion}

\subsection{Limitations}
As with any systematic literature review, this work exhibits some notable limitations related to inclusion and exclusion of relevant work. First, we used the Scopus database as our initial source. Studies have shown Scopus to be an expansive and inclusive source for literature reviews \cite{Falagas2008}, however, a number of important audit studies did not show up in the search and were included through other means. Keywords may have been another factor in these studies not showing up. With the recent growth of "algorithm auditing" as a more defined research area, some initial studies may have used other terms to describe the work, and may have not appeared in our review. Also, our review only included academic papers, thus excluding any audit work published as books (ex. \cite{Noble2018}), journalism (ex. \cite{Kirchner2016}), or other mediums.

\added{The review may also have temporal and linguistic limitations. Many of the papers identified from additional sources (N=36) were published in or before 2015, suggesting that the initial corpus search suffered from a recency bias. This could have excluded older papers that did not self-identify as algorithm audit work at the time of publication. The literature review was also scoped to studies published in English. It may be true that most algorithm audits have come from English-speaking countries, and have focused on algorithmic systems in those countries. In any case, future studies would likely benefit from auditing algorithmic systems in non English-speaking countries.}

\subsubsection{Author Standpoint}
Another limitation of this review is the limited perspective of the author. \added{As with any qualitative work, the researcher's positionality can affect all aspects of a literature review, and it is therefore important to name some characteristics of this position \cite{Harding2004}. The first author only speaks English and has never lived outside the United States. Furthermore, as a person who identifies and passes as a cisgender white male in the United States, he benefits from a number of social systems and structures, including capitalism, colonialism, cis-normativity, whiteness, and patriarchy. The author's privileged standpoint within these systems likely limits how the literature review addresses the oppression and violence inflicted by algorithmic systems.}

\subsection{Toward Algorithmic Justice}
This review has shown how algorithm audits can clarify--and thus help improve--the relationship between technology and society. It is important to recognize how algorithm audits fit into broader work that addresses this relationship and aims to work toward "algorithmic justice," to use a term from \citet{Raji}. In the words of \citet{Abebe2019}, "meaningful advancement toward social change is always the work of many hands." To situate algorithm auditing within a broader context, this subsection aims to identify some of the "many hands" working toward algorithmic justice within the realm of human-computer interaction research. Specifically, we note researchers working on algorithmic justice through user studies, development histories, non-public algorithm audits, and case studies.

\subsubsection{User Studies}
User studies can help uncover problematic algorithm behavior through indirect methods such as user surveys and interviews \cite{Fletcher2018b,Eiband2019a,Lee2015Machines}. While these user studies were not included in the review, they provide an effective method for identifying potentially problematic algorithms and painting the "before-and-after picture," as Nissenbaum calls it \cite{Nissenbaum2001}, of the whole socio-technical system. For example, in interviews with Uber and Lyft drivers, \citet{Lee2015Machines} found that algorithmic assignments often did not make sense to the workers, a finding which could help guide audit studies of the assignment algorithms. In a similar vein, \citet{Diaz2019b} interviewed residents about an algorithmic neighborhood walkability score, finding that many factors the residents named (ex. places of worship and transit access) were missing in the patented Walk Score algorithm.

\subsubsection{Development Histories}
Studies that explore the development history of an algorithmic system provide important narrative frameworks for algorithmic justice work, which can help inform algorithm audits. For example, \citet{Gillingham2016} demonstrates a muddled development of predictive tools for child protection services in New Zealand. \citet{DeVito2017} conducts a content analysis of patents, press releases, and other documents to show how values such as personal preferences were considered in Facebook's design of the News Feed algorithm. Future algorithm audit studies may use these development histories to help select problematic behaviors or organizations worth auditing.

\subsubsection{Non-public Audits}
While this review did not include several audits of non-public algorithms (due to the focus on public accountability), these audits can help characterize potential and/or eminent problems with algorithmic systems. Some of these non-public audits found age-related discrimination in sentiment analysis models \cite{Diaz2018a}, gender discrimination in semantic representation algorithms that may be used in hiring \cite{De-Arteaga2019}, and age-related disparities in landmark detection algorithms often used for facial analysis \cite{Taati2019a}. \citet{Karakasidis2019a} found evidence that name matching algorithms may exhibit racial discrimination, and \citet{Binns2017c} found that models used for language processing can exhibit performance disparities along gendered lines. This early evidence can help inform future audits of public-facing algorithmic systems that rely on or resemble these non-public algorithms.

\subsubsection{Case Studies}
Finally, small-scale audits can have a large impact, \added{as demonstrated by some studies included in this review}. As mentioned in the computer vision section, a single image-tagging result ("Google Mistakenly Tags Black People as `Gorillas'" \cite{Barr2015}) led to wide publicity and swift corporate response. \citet{Gillespie2016} authored one "small-scale" case study that vividly demonstrated the social and political complexity of internet search using just one search term ("santorum"). Another salient example is the paper by \citet{Noble2013}, which exposed significant problems using a single screenshot of a Google search for "black girls." For these empirical examples, small scale is a strength rather than a weakness, as they provide extremely powerful evidence for diagnosing "how [social problems] manifest in technical systems" \cite{Abebe2019}, especially in terms of representational harms \cite{Barocas2017}. As framed by \citet{Benjamin2019Race}, case studies may surface a "glitch" pointing to a broader representational harm, which could then guide justice-oriented changes or inform larger-scale studies.

\section{Conclusion}
This systematic literature review focused on audit studies of public-facing algorithmic systems, synthesizing findings from 62 studies and identifying recurring types of problematic machine behavior. The review shows that algorithm audits have \added{diagnosed} a range of problematic machine behaviors, such as discrimination in advertising algorithms and distortion in search algorithms. \added{These studies provide empirical evidence that the public harms of algorithmic systems are not theoretical conjectures, rather, they play out in real-world public systems affecting millions of people. The review also suggests that some areas are ripe for future algorithm audits, including addressing discrimination in terms of intersectional identities, further exploring advertising algorithms which are the economic backbone of large technology companies, and employing under-explored methods such as code auditing. Some organizations (e.g. Twitter, TikTok, LinkedIn) also deserve further attention. If future audits continue to examine public-facing algorithms, hone in on specific problematic behavior(s), and use compelling baselines, then the field of algorithm auditing can continue holding algorithmic systems accountable by diagnosing harms they pose to the public.}

\begin{acks}
\added{I greatly appreciate Dr. Michelle Shumate for guiding this literature review in her class, "The Practice of Scholarship," and in a subsequent independent study. Thanks also to Daniel Trielli, who helped immensely in synthesizing and coding the papers, and to Priyanka Nanayakkara, who generously reviewed a draft of this paper and provided insightful feedback.}
\end{acks}

\bibliographystyle{ACM-Reference-Format}
\bibliography{library}



\end{document}